# COVID-19 diagnosis by routine blood tests using machine learning

Matjaž Kukar,[1,2*] Gregor Gunčar,[1,3*] Tomaž Vovko,[4] Simon Podnar,[5] Peter Černelč,[7] Miran Brvar,[6] Mateja Zalaznik,[4] Mateja Notar,[1] Sašo Moškon,[1] Marko Notar[#1]

[1]Smart Blood Analytics Swiss SA, CH-8008 Zürich, Switzerland
[2]Faculty of Computer and Information Science, University of Ljubljana, Slovenia
[3]Faculty of Chemistry and Chemical Technology, University of Ljubljana, Slovenia
[4]Department of Infectious Diseases, University Medical Centre Ljubljana, Slovenia
[5]Division of Neurology, University Medical Centre Ljubljana, Slovenia
[6]Centre for Clinical Toxicology and Pharmacology, University Medical Centre Ljubljana, Slovenia
[7]Division of Internal Medicine, University Medical Centre Ljubljana, Slovenia

[*]Joint first authors, contributed equally.

[#]Corresponding author marko@sba-swiss.com

## Abstract

Physicians taking care of patients with coronavirus disease (COVID-19) have described different changes in routine blood parameters. However, these changes, hinder them from performing COVID-19 diagnosis. We constructed a machine learning predictive model for COVID-19 diagnosis. The model was based and cross-validated on the routine blood tests of 5,333 patients with various bacterial and viral infections, and 160 COVID-19-positive patients. We selected operational ROC point at a sensitivity of 81.9% and specificity of 97.9%. The cross-validated area under the curve (AUC) was 0.97. The five most useful routine blood parameters for COVID19 diagnosis according to the feature importance scoring of the XGBoost algorithm were MCHC, eosinophil count, albumin, INR, and prothrombin activity percentage. tSNE visualization showed that the blood parameters of the patients with severe COVID-19 course are more like the parameters of bacterial than viral infection. The reported diagnostic accuracy is at least comparable and probably complementary to RT-PCR and chest CT studies. Patients with fever, cough, myalgia, and other symptoms can now have initial routine blood tests assessed by our diagnostic tool. All patients with a positive COVID-19 prediction would then undergo standard RT-PCR studies to confirm the diagnosis. We believe that our results present a significant contribution to improvements in COVID-19 diagnosis.

## Introduction

In December 2019, cases of pneumonia of an unknown origin were identified in Wuhan, the capital of Hubei province, China [1]. The causative agent was named severe acute respiratory syndrome coronavirus 2 (SARS-CoV-2)[2], and the disease was named coronavirus disease (COVID-19). Soon after, it was realized that SARS-CoV-2 is a highly contagious and moderately virulent virus[3]. In the following months, SARS-CoV-2 spread worldwide, and on March 11, 2020, the World Health Organization (WHO) declared COVID-19 a pandemic [4]. Although clinical features of COVID-19 patients were soon described [5,6], no vaccination or effective treatment was available. Currently, the only effective measures for stopping COVID-19 spread are strict precautionary hygiene, social distancing, and isolation of contagious subjects [7,8].

COVID-19 diagnosis is crucial for the identification, isolation, and treatment of contagious subjects [9]. The gold standard for COVID-19 diagnosis is a demonstration of SARS-CoV-2 RNA in patients' respiratory secretions using real-time reverse transcriptase polymerase chain reaction (RT-PCR) [10,11]. Although RT-PCR is invaluable in dealing with the COVID-19 pandemic, it is a sophisticated test that requires extensive and delicate infrastructure [10]. Moreover, the test is not always positive even in fully symptomatic SARS-CoV-2 infected patients [12]. Some authors have reported only 30%-60% sensitivity of RT-PCR in clinical applications [13,14]. Additionally, demand for RT-PCR testing is enormous, which is a limitation in the control of the pandemic [15]. In symptomatic COVID-19 patients, a CT scan of the chest is a useful [13] but undesirable alternative [16]. Therefore, other testing methods are imperative.

Physicians taking care of COVID-19 patients have noted pronounced changes in their blood parameters. Particularly, they have described hypoalbuminemia, increased C-reactive protein (CRP) and lactate dehydrogenase (LDH), lymphopenia, etc. [17]. Nevertheless, these laboratory findings alone are insufficient for physicians to differentiate patients with COVID-19 from patients with other infectious disorders. More so, it is widely known that even the most knowledgeable and experienced physicians can extract only a minor fraction of information contained in the results of routine blood tests [18]. By contrast, machine learning (ML) can recognize subtle patterns in data. Therefore, ML is suitable for differentiating various patterns observed in routine blood parameters. We have previously demonstrated how an ML model considerably outperformed experienced clinicians in diagnosing hematological disorder [18], as well as another model (similar to head imaging) with diagnostic accuracy for brain tumors [19].

The aim of the present study is to determine the diagnostic accuracy of a ML model built specifically for the diagnosis of COVID-19 using results of routine blood tests. A group of symptomatic patients newly diagnosed with COVID-19 and patients with other infectious diseases were studied.

## Materials and methods

### Patients and controls

A pool of a COVID-19-positive population was obtained in March/April 2020 from patients admitted to the Department of Infectious Diseases, University Medical Centre Ljubljana (UMCL), Slovenia. The training positive group included 160 consecutive symptomatic patients.

A pool of a COVID-19-negative population was obtained from 52,306 patients admitted to the same Department from March, 2012 to April, 2019. A more representative population of 22,385 patients with various viral and bacterial infections, and approximately the same mean number of measured blood parameters as in the COVID-19-positive patients (33 out of 35) was selected. To construct the final representative training negative group, patients were randomly sampled (without replacement) to approximate the proportion of positive versus tested individuals (3% at the time of data collection). At the end, the training negative group included retrospective data of 5,333 patients with 225 different bacterial and viral infections (different ICD codes), diagnosed prior to the COVID-19 outbreak (Fig 1).

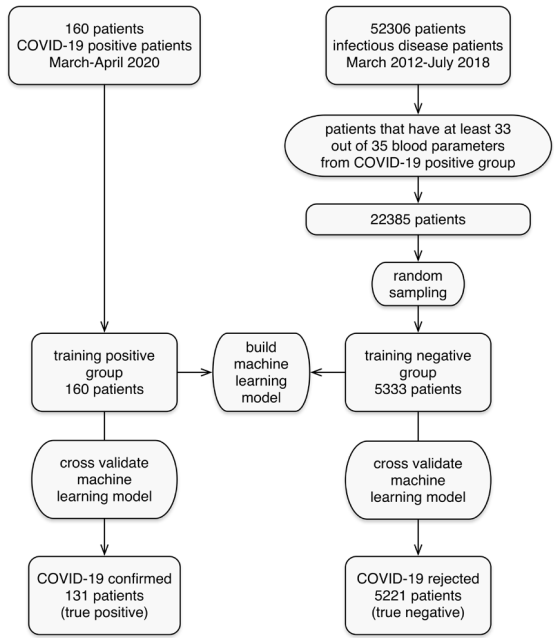

**Fig 1. A flow chart of patients included in the model building and validation process.**

In all groups, we collected data on patients' age, sex, routine blood test results, and the ICD10-encoded final diagnoses. All identifiable personal data items were removed prior to analysis. The National Ethics Committee of Slovenia approved the study (No. 0120-718/2015/7, No. 0120-170/2020/6). Due to the retrospective design, the necessity of obtaining patients' written informed consent was waived. The study was performed in accordance with the STARD recommendations [20].

## Blood parameters used for model building

Out of 117 parameters measured in the training positive group, we removed all parameters that were measured in less than 25% of the patients. We also omitted non-blood parameters and arterial blood parameters. Thus, 35 parameters were selected. For each parameter, we calculated the relative reference range and median values for a group of patients with COVID-19, and in the training negative group, we calculated for the viral and bacterial infections separately. All parameter values (reference ranges, medians) were centered and scaled according to reference ranges. We compared blood parameter distributions in groups by the nonparametric k-sample Anderson-Darling (AD) test and depicted the *P*-values [21].

## Visualization of blood parameter space

To visualize how the data was arranged in a high-dimensional space of 35 blood parameters, we applied the t-distributed stochastic neighbor embedding (t-SNE) method [22], which is an unsupervised, non-linear technique primarily used for data exploration and visualization of high-dimensional data. The method has been shown to perform effectively in several high-dimensional datasets, it is very flexible, and it can often find a structure where other dimensionality-reduction algorithms fail [22,23]. The nature and complexity of t-SNE may lead to visualization misinterpretation, specifically to overstating the meaning of distances on the plot [24]. In this work, we used the openTSNE implementation [25,26].

### Smart Blood Analytics machine learning algorithm

The Smart Blood Analytics (SBA) algorithm is a CRISP-DM based machine learning pipeline consisting of five processing stages corresponding to phases 2-6 of the CRISP-DM [27] standard. The stages are as follows. Data acquisition: acquiring raw data from the database; data filtering: constructing the training dataset consisting of blood test results obtained before treatment and the patient's final diagnosis; data preprocessing: canonization of blood parameters (matching them with our reference blood parameter database, recalculation to SI units, data quality control); data modelling: building the diagnostic model using ML algorithms; evaluation: evaluating the model with stratified ten-fold cross-validation and/or independent testing data; deployment of the successfully evaluated model in the cloud (accessible either through hospital information systems or the SBA website[28]). As the principal ML algorithm, we utilized either random forests (RF), deep neural networks (DNN), or the extreme gradient boosting machine (XGBoost). To diagnose the COVID-19 disease, the XGBoost [29-31] learning algorithm was chosen after initial experimental runs of all three candidates as it produced models with significantly higher performances and required less computational resources. Incidentally, XGBoost is currently one of the most popular ML tools [32] with key strengths, such as speed and parallelization; more importantly, it can intrinsically handle sparse (missing) data, which many other algorithms struggle with [33].

### Imbalanced data and model calibration

In our data, we observed severely imbalanced groups (in daily practice, the ratio of positive versus tested is approximately 3%). However, such a scenario is often problematic for machine learning algorithms as it makes it too easy to focus on the prevalent group (negatives). Simple data undersampling techniques failed to improve the results due to the relatively large number of blood parameters and a correspondingly large (35+2)-dimensional attribute space. Moreover, more advanced resampling techniques, such as SMOTE [34,35], struggle with high-dimensional and interdependent data [36], such as blood test measurements. Therefore, undersampling was only used to maintain the ratio between the positive and negative groups. Additionally, the intrinsic imbalance was addressed by model calibration using the precision-recall (PR) curve [37] and maximizing the F2-score (favoring recall versus precision) to select the operational ROC point.

### Evaluation of predictive models

The models were evaluated in two different ways. First, we automatically evaluated the models using repeated stratified ten-fold cross-validation. The results were characterized using standard performance measures, such as sensitivity and specificity (recall on positive and negative groups, respectively), precision, AUC, and ROC curve. Additionally, we tested the final model on a separate control group and reported the same performance measures. Furthermore, for sensitivity and specificity, the 95% binominal confidence intervals using the Agresti-Coull method were calculated [38].

## Results

Demographic data for all patient groups are presented in table 1. Out of the 160 COVID-19-positive patients (median age: 55.5 years; 42% women), 17 were admitted to the intensive care unit (ICU), and 14 required intubation and invasive mechanical ventilation. Chest X-ray was performed on 94 patients, and lung infiltrates were detected in 68 patients. Respira-tory failure occurred in 44 patients (27.5%), 10 died (6%), 7 were still in the ICU (4%), and 20 were in the hospital (12.5%). The following comorbidities were also present: hypertension in 34.4%, diabetes in 9.4%, hyperlipidemia in 11.9%, heart failure in 7.5%, hypothyroidismin in 6.3%, atrial fibrillation in 5.0%, ischemic heart disease in 3.8%, COPD or asthma in 5.6%, chronic kidney failure in 3.8%, and occlusive peripheral arterial disease in 1.9%.

Table 1. Demographic features of included patient groups.

|  | Training group – COVID-19 | |
| --- | --- | --- |
|  | Negative | Positive |
| **Number** | 5333 | 160 |
| **Age median** | 57 | 55.5 |
| **Female number/%** | 2155/40% | 67/42% |

The analysis of 35 selected blood parameters revealed that in the COVID-19 positive group, the calculated parameter medians were within the normal reference range for all except two parameters that were elevated: prothrombin activity % (median: 1.05; normal range (SI): 0.7-1), and CRP (median: 12 mg/L; SI: 0-5 mg/L) (Fig 2). Most blood test parameters from the patients with COVID-19 differed significantly from patients with other viral and bacterial infections. Five parameters with the statistically most significant difference and effect size between the COVID-19-positive group and bacterial infections were urea, hemoglobin, erythrocyte count, hematocrit, and leukocyte count. When the COVID-19-positive group was compared to other viral infections, the five parameters with the statistically most significant difference and effect size were mean corpuscular hemoglobin concentration (MCHC), eosinophils ratio, prothrombin international normalized ratio (INR), prothrombin activity %, and creatinine (Fig 2). According to the XGBoost feature importance scoring, the five blood parameters with the highest discriminative power were MCHC, eosinophils count, albumin, INR and prothrombin activity %.

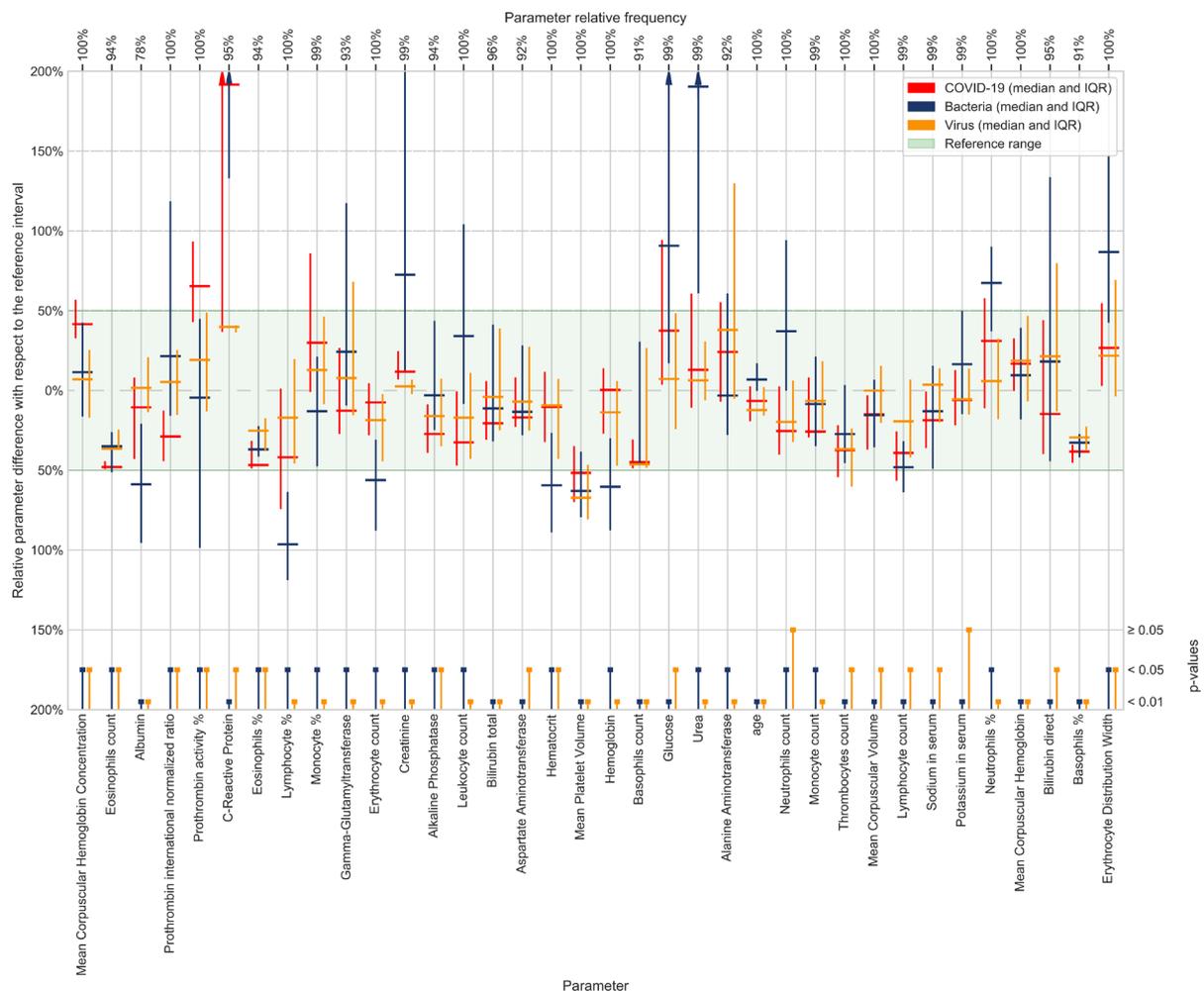

**Fig 2. Blood parameters** sorted by their XGBoost importance score (more important parameters are shown on the left). Group median values and IQR of the blood parameters used in model building are shown, centered, and scaled to reference intervals. Median bar for the C-reactive protein in bacterial infections is out of the scale at 38 mg/L. Groups (COVID-19/other virus/bacteria) were evaluated by the Anderson-Darling test. The significance levels (0.05 or 0.01) of the test results are depicted at the bottom.

The full complexity of COVID-19 diagnostics can be illustrated by visualizing the blood parameter space of patients with COVID-19, bacterial, and viral infection from our training data using the t-SNE method [22] (Fig 3). Even after extensive experimentation, which also included alternative visualization techniques, such as PCA and MDS, it was impossible to obtain partial separation of the positive and negative groups. While the virus and bacteria subgroups appear different, but have a significant overlap, the COVID-19 positive group is dispersed between both. Expectedly, the medoid of the COVID-19 positive group lies closer to the medoid of the virus subgroup than to the medoid of the bacteria subgroup. This is not the case in the COVID-19 positive patients who died or had a diagnosis of acute respiratory failure (ARF). The medoids of those patients are both closer to the medoid of the bacteria subgroup (Fig 3).

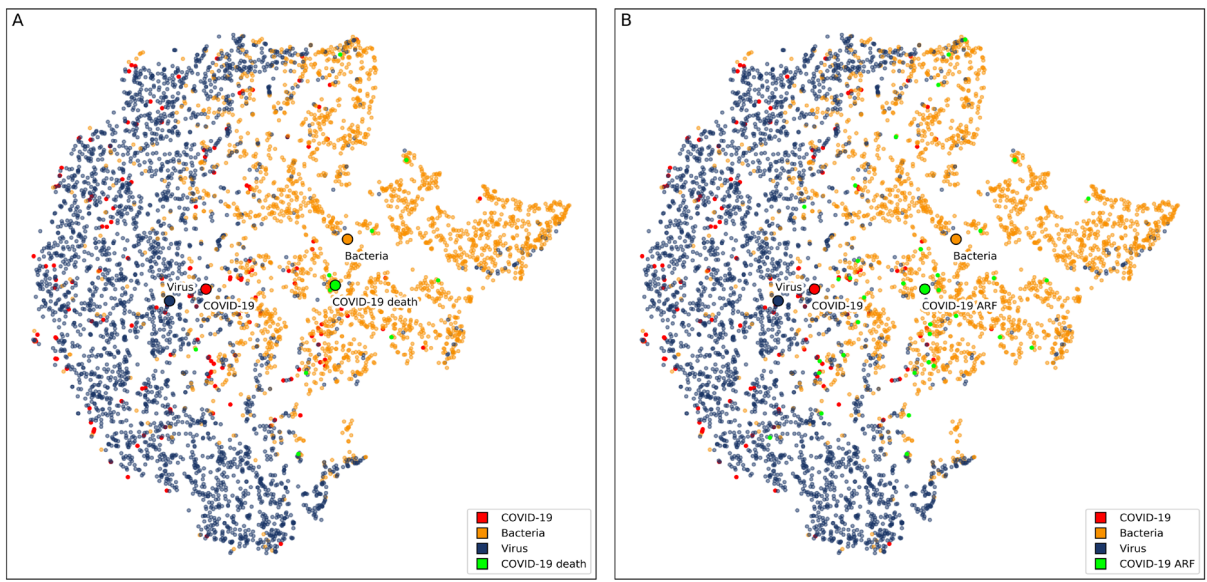

**Fig 3. Visualization of bacteria/virus/COVID-19 parameter space with t-SNE method.** Medoids of bacteria/virus/COVID-19 groups are marked. Each dot represents a patient (more specifically, an embedding of his/her blood parameters into a two-dimensional space), and its color represents the group. Green dots represent patients who died (A) and patients diagnosed with acute respiratory failure (B).

Nevertheless, the predictive model for the diagnosis of COVID-19, which was produced using XGBoost, performed effectively (Fig 2). In the training group, using ten-fold stratified cross-validation testing procedure, the results and the corresponding binomial confidence intervals, calibrated with respect to the operational ROC point were as follows: a sensitivity of 81.9% ± 6, specificity of 97.9% ± 0.4%, and AUC of 0.97 (table 2, Fig 4).

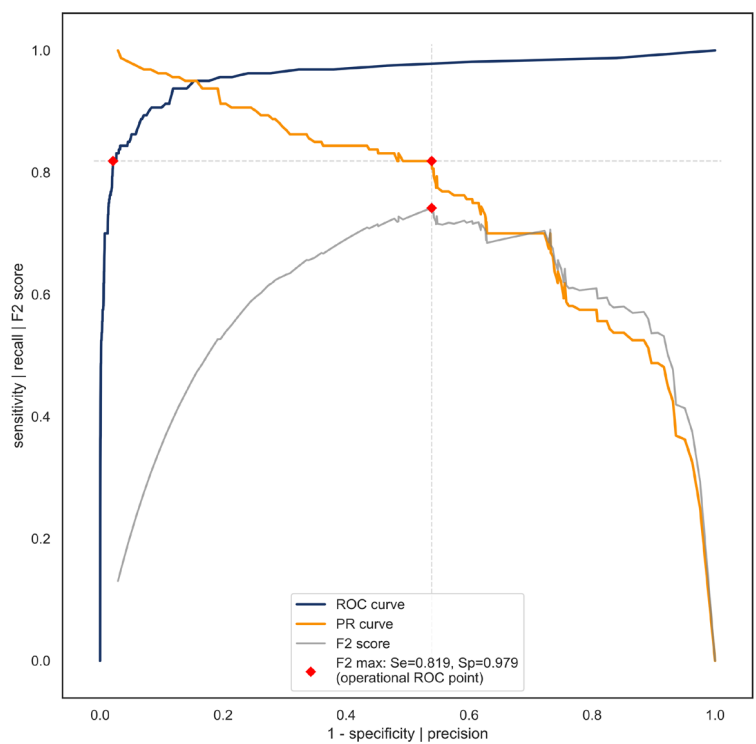

**Fig 4. ROC, PR (precision-recall), and F2 curves for COVID-19 diagnosis calculated from the training data using ten-fold stratified cross-validation.** Vertical and horizontal dashed lines connect the F2 (gray) max point with the PR curve (orange) and the ROC curve (blue) in order to obtain the operational ROC point with sensitivity = 0.819, specificity = 0.979 (depicted with red dots), and AUC = 0.97.

**Table 2. Confusion table for the cross-validated training group**

|  | Positive | Negative |
|---|---|---|
| **Predicted positive** | 131 | 112 |
| **Predicted negative** | 29 | 5221 |

# Discussion

In this study, we confirmed that COVID-19 diagnosis is attainable using ML on data from routine blood tests. We demonstrated that our ML model efficiently discriminated patients with COVID-19 from patients with other infectious diseases. The model exhibited a high sensitivity of 81.9%, a specificity of 97.9%, and an AUC of 0.97 on the cross-validated training group (Fig 4). From an ML perspective, our results are quantitatively excellent, with an impressively low proportion of false positives and a moderately low proportion of false negatives. Moreover, AUC values above 0.90 are generally considered as excellent [39].

Owing to the absence of a completely reliable diagnostic standard for COVID-19, it is difficult to evaluate the diagnostic performance of various diagnostic tests. Nevertheless, it is clear that the diagnostic performances of both RT-PCR studies and chest CT are not perfect. In a recent study of 1,014 patients suspected with COVID-19, both tests were positive in 580 cases, only chest CT was positive in 308, only RT-PCR in 21, and none of them in the remaining 105 patients; RT-PCR sensitivity was 59%, and chest CT was 88% [13]. The diagnostic performance of our predictive model is most likely not inferior to its competitors. Furthermore, it is most probably complementary and would be best used along with standard protocols designed according to local circumstances.

In a study describing a ML model using blood parameters [40], the researchers studied 105 patients with COVID-19 and 148 patients with other pulmonary disorders. They identified 11 most useful blood parameters (total protein, bilirubin, glucose, creatinine, Ca, LDH, creatine kinase, K, Mg, platelet distribution width, and basophil count) and used them in their analyses. They also recorded high test accuracies: 98% on cross-validation and 97% on test set [40]. Although their work has not been peer-reviewed and published in a scientific literature, their data confirm our finding that ML models using routine blood parameters is useful in the diagnosis of COVID-19. However, their data quantitatively has a 41% ratio of positives. Thus, where the ratio is much lower in practice, unacceptably high numbers of false positives would be recorded.

In another study, the authors collected 102 patients diagnosed as positive and 133 diagnosed as negative with RT-PCR tests.[41] Their best results are considerably lower than ours (AUC: 0.85, sensitivity 0.68, specificity 0.85), most likely due to a much lower number of blood parameters measured (only 13). As demonstrated in the study [40], it is difficult to assess the practical importance of their results as the 43% ratio of positives would in practice be much smaller and again result in high numbers of false positives.

We obtained blood samples from our patients immediately after they were presented to the infectious disease service. This observation suggests that the SBA algorithm is useful in the early symptomatic phase when COVID-19 is easier to be missed by RT-PCR test. We do not have data on the ability of our model to diagnose presymptomatic COVID-19 patients as their blood had not been drawn. Although this should be tested in the future, our model will possibly be inefficient at that stage in which the virus replicates locally in the nasopharynx without systemic effects.

Some routine blood parameters proved to be especially important in our model. It should be noted that we selected the blood parameters we used for model training and analysis based on the available data in all of our patient groups. Therefore, we were unable to include some clinically relevant parameters that might be helpful in identifying patients with COVID-19. However, our analysis revealed some blood parameters that require further investigation in patients with COVID-19. In our analysis, the two out of five most discriminating parameters for patients with COVID-19 were prothrombin activity % and INR, which were elevated and decreased, respectively, indicating accelerated blood clot formation in patients with COVID-19. The risk of disseminated intravascular coagulation and venous thromboembolism is well recognized in COVID-19 [42]. We also observed raised MCHC, a reduction in eosinophils, low albumin levels, high CRP, and lymphopenia (Fig 2). In a systematic review and meta-analysis of 19 studies, the most prevalent laboratory abnormalities found in patients with COVID-19 were hypoalbuminemia (76%), increased CRP (58%), LDH (57%), and lymphopenia (43%) [17]. However, this pattern of abnormalities is still rather nonspecific and does not enable physicians to diagnose COVID-19. Likewise, considering the 35 most important parameters we analyzed (Fig 2) does not enable physicians to confirm COVID-19 diagnosis. This is also evident from our t-SNE analysis and visualization of the distribution of COVID-19, bacterial infection, and viral infection cases, which showed the complexity of the parameter space in COVID-19 (Fig 3). Apart from diagnosis, physicians caring for patients with COVID-19 also noted some typical patterns in blood pa-rameters that predict more severe disease courses. Most notably in patients with more severe disease courses, laboratory abnormalities were more pronounced (e.g., more severe lymphopenia, CRP and LDH increase, etc.) [5]. In agreement, our t-SNE visualization of blood parameter space shows that the medoid of the patients with severe COVID-19 course is shifted toward the medoid of the patients with bacterial infection (Fig 3). This indicates for the need of COVID-19 patients to be tested for bacteria or severe inflammation early on and treated accordingly. It also shows the possibility of the efficient prognostication of the COVID-19 course using ML.

Our study has several limitations. First, our analysis was performed on data obtained in a single center. Although this may limit generalizability, using standardized and approved procedures, reagents, and technology, we expect similar laboratory blood test results in other centers. Second, the number of COVID-19-positive patients included in our analyses was limited (160 for the building of the ML model). Both data disproportion and parameter dimensionality suggest that a considerably higher number of positive patients (at least 1,000) would further improve results on the positive group. However, with respect to the small number of available COVID-19-positive patients, the current results are excellent. Third, the study was retrospective, which limited the scope of available patient data. However, for the purpose of this study, we mainly required available results of routine blood tests and accurate COVID-19 diagnoses.

The study also has several strengths. First, we analyzed data from a large number of patients (> 5,000) with good data quality for blood tests and diagnoses. Second, a single certified laboratory diagnosed all patients with COVID-19 using RT-PCR, which assured the high quality of the diagnoses. The specificity of RT-PCR was also very high. Furthermore, high specificity was assured by the inclusion of patients evaluated for various infectious diseases before the COVID-19 pandemic. Third, we used state-of-the-art ML algorithms that can develop the best predictive models.

The study demonstrates that symptomatic patients with COVID-19 can be efficiently diagnosed from the results of routine blood tests. The SBA COVID-19 ML model extracted subtle prognostic data from blood test results that were hidden from the most experienced clinicians. We believe that our results present an important step to a more widely available diagnosis of patients with COVID-19. Moreover, our ML predictive model is available worldwide at www.smartbloodanalytics.com as a web application or through an API call, and it can be used instantly. The model will also be of benefit after the pandemic as it will be an alternative for a physician to test patients for COVID-19 from the blood test results of other diagnoses.

## Acknowledgments

We would like to thank Editage for English language editing.